\documentstyle[12pt,epsf,colordvi]{article}
\newcommand{\FRENCH}{0}  % 1 puts abstract/title page in French
\newcommand{\FOOTNOTE}{1}  % 1 includes change of value footnote
\newcommand{\nub}{\overline{\nu}}
\newcommand{\nunub}{\stackrel{{\footnotesize (-)}}{\nu}}
\newcommand{\txnunub}[1]{\nu_{#1}/\nub_{#1}}
\newcommand{\ubar}{\overline{u}}	
\newcommand{\dbar}{\overline{d}}
\newcommand{\qbar}{\overline{q}}

\newcommand{\stw}{\mbox{$\sin^2\theta_W$}}

\newcommand{\rmt}{\rm\textstyle}
\newcommand{\rms}{\rm\scriptstyle}
\newcommand{\nunuc}{\mbox{$\nu$N}}
\newcommand{\rmeas}[1]{\mbox{$R_{\rms meas}^{#1}$}}
\newcommand{\rmeasm}{R_{\rms meas}}
\newcommand{\mw}{\mbox{$M_W$}}

\newcommand{\mwmw}{\mbox{$M_W^2$}}
\newcommand{\mzmz}{\mbox{$M_Z^2$}}
\newcommand{\mtop}{\mbox{$M_{\rms top}$}}
\newcommand{\mt}{\mbox{$M_{\rms top}$}}
\newcommand{\mhiggs}{\mbox{$M_{\rms Higgs}$}}
\newcommand{\compressedsection}[1]{{\section{{#1}}}}
\def\be{\begin{equation}}
\def\ee{\end{equation}}
\def\bea{\begin{eqnarray}}
\def\eea{\end{eqnarray}}
\def\beas{\begin{eqnarray*}}
\def\eeas{\end{eqnarray*}}
\def\err#1#2#3  {{\it Erratum} {\bf#1}, #3 (#2) }
\def\ib#1#2#3   {{\it ibid.} {\bf#1}, #3 (#2) }
\def\ijmp#1#2#3 {{\em Int. J. Mod. Phys.} {\bf#1}, #3 (#2) }
\def\jetp#1#2#3 {{\em JETP Lett.} {\bf#1}, #3 (#2) }
\def\mpl#1#2#3  {{\em Mod. Phys. Lett.} {\bf#1}, #3 (#2) }
\def\nat#1#2#3  {{\em Nature (London)} {\bf#1}, #3 (#2) }
\def\nc#1#2#3   {{\em Nuovo Cim.} {\bf#1}, #3 (#2) }
\def\nim#1#2#3  {{\em Nucl. Instr. Meth.} {\bf#1}, #3 (#2) }
\def\np#1#2#3   {{\em Nucl. Phys.} {\bf#1}, #3 (#2) }
\def\pcps#1#2#3 {{\em Proc. Cam. Phil. Soc.} {\bf#1}, #3 (#2) }
\def\pl#1#2#3   {{\em Phys. Lett.} {\bf#1}, #3 (#2) }
\def\prep#1#2#3 {{\em Phys. Rep.} {\bf#1}, #3 (#2) }
\def\prev#1#2#3 {{\em Phys. Rev.} {\bf#1}, #3 (#2) }
\def\prl#1#2#3  {{\em Phys. Rev. Lett.} {\bf#1}, #3 (#2) }
\def\prs#1#2#3  {{\em Proc. Roy. Soc.} {\bf#1}, #3 (#2) }
\def\ptp#1#2#3  {{\em Prog. Th. Phys.} {\bf#1}, #3 (#2) }
\def\rmp#1#2#3  {{\em Rev. Mod. Phys.} {\bf#1}, #3 (#2) }
\def\rpp#1#2#3  {{\em Rep. Prog. Phys.} {\bf#1}, #3 (#2) }
\def\sjnp#1#2#3 {{\em Sov. J. Nucl. Phys.} {\bf#1}, #3 (#2) }
\def\spj#1#2#3  {{\em Sov. Phys. JEPT} {\bf#1}, #3 (#2) }
\def\zp#1#2#3   {{\em Zeit. Phys.} {\bf#1}, #3 (#2) }
\def\epj#1#2#3   {{\em Eur. Phys. Jour.} {\bf#1}, #3 (#2) }
\begin{document}

\newcommand{\linespace}[1]{\protect\renewcommand{\baselinestretch}{#1}
  \footnotesize\normalsize}
\begin{flushright} 
\typeout{need preprint number} % preprint number here
\end{flushright}
\begin{center} 
{\ifnum\FRENCH=1
  {\large LA MEASURE DE \stw\ DANS LA DIFFUSION PROFONDE INELASTIQUE
    $\nu-N$ A L'EXPERIENCE NUTEV}
\else 
  {\large MEASUREMENT OF \stw\ FROM NEUTRINO-NUCLEON SCATTERING AT NuTeV}
\fi}
\end{center} 
{\footnotesize
\begin{center}
\begin{sloppypar}
\noindent
 K.~S.~McFarland$^{3,\dagger}$, T.~Adams$^{4}$, A.~Alton$^{4}$,
 S.~Avvakumov$^{7}$, L.~de~Barbaro$^{5}$, P.~de~Barbaro$^{7}$,
 R.~H.~Bernstein$^{3}$, A.~Bodek$^{7}$, T.~Bolton$^{4}$,
 J.~Brau$^{6}$, D.~Buchholz$^{5}$, H.~Budd$^{7}$, L.~Bugel$^{3}$,
 J.~Conrad$^{2}$, R.~B.~Drucker$^{6}$, R.~Frey$^{6}$,
 J.~Goldman$^{4}$, M.~Goncharov$^{4}$, D.~A.~Harris$^{7}$,
 R.~A.~Johnson$^{1}$, S.~Koutsoliotas$^{2,*}$, J.~H.~Kim$^{2}$,
 M.~J.~Lamm$^{3}$, W.~Marsh$^{3}$, D.~Mason$^{6}$, C.~McNulty$^{2}$,
 D.~Naples$^{4}$, P.~Nienaber$^{3}$, A.~Romosan$^{2,\ddagger}$,
 W.~K.~Sakumoto$^{7}$, H.~Schellman$^{5}$, M.~H.~Shaevitz$^{2}$,
 P.~Spentzouris$^{3}$ , E.~G.~Stern$^{2}$, M.~Vakili$^{1,\parallel}$,
 A.~Vaitaitis$^{2}$, V.~Wu$^{1}$, U.~K.~Yang$^{7}$, J.~Yu$^{3}$ and
 G.~P.~Zeller$^{5}$
\vspace{15pt} 

 $^{1}$University of Cincinnati, Cincinnati, OH 45221 USA \\            %1
 $^{2}$Columbia University, New York, NY 10027 USA \\                   %2
 $^{3}$Fermi National Accelerator Laboratory, Batavia, IL 60510 USA \\  %3
 $^{4}$Kansas State University, Manhattan, KS 66506 USA \\              %4
 $^{5}$Northwestern University, Evanston, IL 60208 USA \\               %5
 $^{6}$University of Oregon, Eugene, OR 97403 USA \\                    %6
 $^{7}$University of Rochester, Rochester, NY 14627 USA \\              %7
\vspace{5pt}

{\ifnum\FRENCH=1
 $^{\dagger}$ Adresse actuel: Massachusetts Institute of Technology,
                  Cambridge, MA 02139 USA \\
 $^{*}$ Adresse actuel: Bucknell University, Lewisburg, PA 17837 USA \\
 $^{\ddagger}$ Adresse actuel: Lawrence Berkeley National Laboratory, 
                      Berkeley, CA 94720 USA \\
 $^{\parallel}$ Adresse acutel: Texas A\&M University, 
                       College Station, TX 77801 USA
\else 
 $^{\dagger}$ Current address: Massachusetts Institute of Technology,
                  Cambridge, MA 02139 USA \\
 $^{*}$ Current address: Bucknell University, Lewisburg, PA 17837 USA \\
 $^{\ddagger}$ Current address: Lawrence Berkeley National Laboratory, 
                      Berkeley, CA 94720 USA \\
 $^{\parallel}$ Current address: Texas A\&M University, 
                       College Station, TX 77801 USA
\fi}

\end{sloppypar}
\end{center}
}
\vspace{1cm} 

\begin{center}
Submitted to the Proceedings of the XXXIIIrd Recontres de Moriond \\
March 1998, Les Arcs, France
\end{center}
\begin{abstract} 
{\ifnum\FRENCH=1
Nous r\'{e}portons la measure de \stw\ dans la diffusion 
profonde inelastique $\nu-N$ de l'exp\'{e}rience NuTeV.  
Utilisant les faisceaux separ\'{e}s entre neutrino et 
anti-neutrino, NuTeV peut d\'{e}terminer la variable de 
Paschos-Wolfenstein $R^-=(\sigma^\nu_{\rms NC}-\sigma^{\nub}_{\rms NC})/
(\sigma^\nu_{\rms CC}-\sigma^{\nub}_{\rms CC})$.
NuTeV mesure 
\else
We report the measurement of \stw\ in $\nu-N$ deep inelastic
scattering from the NuTeV experiment.  Using separate neutrino and
anti-neutrino beams, NuTeV is able to determine \stw\ with low
systematic errors by measuring the Paschos-Wolfenstein variable $R^-$,
a ratio of differences of neutrino and anti-neutrino neutral-current
and charged-current cross-sections.  NuTeV measures 
\fi}
\stw$^{({\rms on-shell)}}=0.2253\pm0.0019({\rmt stat})\pm0.0010({\rmt syst})$, 
{\ifnum\FRENCH=1
qui indique une measure de la masse de W,
\else
which implies 
\fi}
\mw$=80.26\pm0.11~$GeV.
\end{abstract} 

\pagebreak

\compressedsection{Introduction}

In the past, neutrino scattering experiments have played a key role in
establishing the validity of the electroweak Standard Model.  Today,
even with the large samples of on-shell $W$ and $Z$ bosons at $e^+e^-$
and $p\overline{p}$ colliders, precision measurements of
neutrino-nucleon scattering still play an important role.  Not only
are these measurements competitive in precision with direct probes of
weak boson parameters, but they also test the validity of the
electroweak theory in different processes and over many orders of
magnitude in $q^2$.  In this respect, if neutrino scattering observed
deviations from expectations based on direct measurements from $W$ and
$Z$ bosons, this would be an exciting hint of new physics entering in
tree-level processes or in radiative corrections.  In particular,
neutrino scattering would be sensitive to a different menu of
non-Standard Model effects ranging from leptoquark exchange to
neutrino oscillations\cite{langacker,oscpaper}.

Experimental quantities sensitive to electroweak physics that
are most precisely measured in neutrino scattering are the ratios
of charged-current ($W$ exchange) to neutral-current ($Z$ exchange)
scattering cross-sections from quarks in heavy nuclei.  The ratio of
these cross-sections for either neutrino or anti-neutrino scattering
from isoscalar targets of $u$ and $d$ quarks can be written as\cite{llewellyn}
\begin{equation}
R^{\nu(\nub)} \equiv \frac{\sigma(\nunub_{\mu}N\rightarrow\nunub_{\mu}X)}
                 {\sigma(\nunub_{\mu}N\rightarrow\mu^{-(+)}X)}  
= (g_L^2+r^{(-1)}g_R^2),
\label{eqn:ls}
\end{equation}
where
\begin{equation}
r \equiv \frac{\sigma({\overline \nu}_{\mu}N\rightarrow\mu^+X)}
                {\sigma(\nu_{\mu}N\rightarrow\mu^-X)} \sim \frac{1}{2},  
\label{eqn:rdef} 
\end{equation}
and $g_{L,R}^2=u_{L,R}^2+d_{L,R}^2$, the isoscalar sums of the squared
left or right-handed quark couplings to the $Z$.  At tree level in the
Standard Model, $q_L=I^{(3)}_{\rms weak}-Q_{\rms EM}\stw$ and 
$q_R=-Q_{\rms EM}\stw$; therefore, $R^\nu$ is particularly sensitive
to \stw.

In a real target, there are corrections to Eqn.~\ref{eqn:ls} resulting
from the presence of heavy quarks in the sea, the production of heavy
quarks in the target, non leading-order quark-parton model terms in
the cross-section, electromagnetic radiative corrections and any
isovector component of the light quarks in the target.  In particular,
in the case where a charm-quark is produced from scattering off of
low-$x$ sea quarks, the uncertainties resulting from the effective
mass suppression of the heavy final-state charm quark are large.  The
uncertainty in this suppression
ultimately limited the precision of previous $\nu N$ scattering
experiments which measured electroweak
parameters\cite{CCFR,CDHS,CHARM}.

To eliminate the effect of uncertainties resulting from scattering
from sea quarks, one can instead form a quantity suggested by Paschos
and Wolfenstein\cite{Paschos-Wolfenstein}, 
\begin{equation}
R^{-} \equiv \frac{\sigma(\nu_{\mu}N\rightarrow\nu_{\mu}X)-
                   \sigma(\nub_{\mu}N\rightarrow\nub_{\mu}X)}
                  {\sigma(\nu_{\mu}N\rightarrow\mu^-X)-  
                   \sigma(\nub_{\mu}N\rightarrow\mu^+X)}  
= \frac{R^{\nu}-rR^{\nub}}{1-r}=(g_L^2-g_R^2).
\label{eqn:rminus}
\end{equation}
Since $\sigma^{\nu q}=\sigma^{\nub\, \qbar}$ and 
 $\sigma^{\nub q}=\sigma^{\nu \qbar}$, the effect of scattering
from sea quarks, which is symmetric under $q\leftrightarrow\qbar$, cancels
in the difference of neutrino and anti-neutrino cross-sections.  While
allowing substantially reduced uncertainties, $R^-$ is a more difficult
quantity to measure than $R^\nu$, primarily because neutral current
neutrino and anti-neutrino scattering have identical observed final
states and can only be separated by {\em a priori} knowledge of the
initial state neutrino.

\compressedsection{The NuTeV Experiment and Neutrino Beam}

The NuTeV detector consists of an $18$~m long, $690$~ton target
calorimeter with a mean density of $4.2$~g/cm$^3$, followed by an iron
toroid spectrometer.  The target calorimeter consists of 168 iron
plates, $3$m~$\times$~$3$m~$\times$~$5.1$cm each.  The active elements
are liquid scintillation counters spaced every two plates and drift
chambers spaced every four plates.  There are a total of 84
scintillation counters and 42 drift chambers in the target. The toroid
spectrometer is not directly used in this analysis.  NuTeV used a
continuous test beam of hadrons, muons and electrons to calibrate the
calorimeter and toroid response.

In this detector $\txnunub{\mu}$ charged-current events are identified
by the presence of an energetic muon in the final state which travels
a long distance in the target calorimeter.  Quantitatively, a length
is measured for each event based on the number of neighboring
scintillation counters above a low threshold.  Charged-current
candidates are those events with a length of greater than $20$
counters ($2.1$~m of steel-equivalent), and all other events are
neutral-current candidates.

NuTeV's target calorimeter sits in the Sign-Selected Quadrupole Train
(SSQT) neutrino beam at the FNAL TeVatron.  The observed neutrinos
result from decays of pions and kaons produced from the interactions
of $800$~GeV protons in a production target.  Immediately downstream
of the target, a dipole magnet with $\int Bdl=5.2$~T-m bends pions
and kaons of one charge in the direction of the NuTeV detector, while
oppositely charged and neutral mesons are stopped in dumps.
Focusing magnets then direct the sign-selected mesons into a $0.5$~km
decay region which ends $0.9$~km upstream of the NuTeV detector.  The
resulting beam is either almost purely neutrino or anti-neutrino,
depending of the selected sign of mesons.  Anti-particle
backgrounds are observed at a level of less than 1--2 parts in $10^3$.
The beam is almost entirely muon neutrinos, with electron neutrinos
creating $1.3\%$ and $1.1\%$ of the observed interactions from the
neutrino and anti-neutrino beams, respectively.

\begin{figure}[tbp]
\epsfxsize=\textwidth\epsfbox{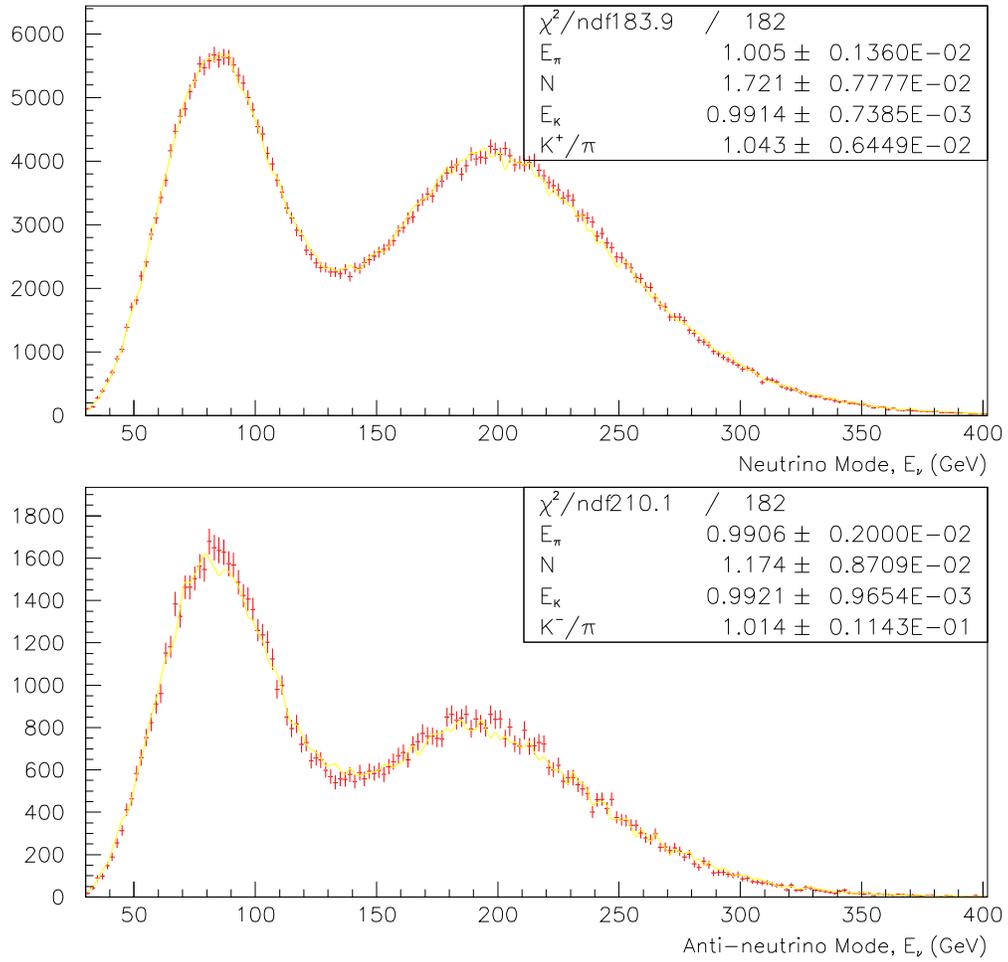}\vspace*{-3ex}
\caption{The $\nu_\mu$ and $\nub_\mu$ energy spectra from the
data and the tuned beam Monte Carlo.}
\label{fig:flux-tune}
\end{figure}

Because charged-current electron neutrino interactions usually lack an
energetic muon in the final state, they are almost always identified
as neutral-current interactions in the NuTeV detector.  Therefore, the
electron neutrino content of the beam must be very precisely known.
Most ($93\%$ in the neutrino beam and $70\%$ in the anti-neutrino
beam) observed $\txnunub{e}$s result from $K^\pm_{e3}$ decays.  The
remainder are products of prompt decays of charmed particles or
neutral kaons, or decays of secondary muons.  Prediction of the former
component comes from a beam Monte Carlo, tuned to reproduce the
observed $\txnunub{\mu}$ flux (Figure~\ref{fig:flux-tune}).  Because
of the precise alignment of the magnetic optics in the SSQT, this
procedure results in a fractional uncertainty on the prediction of
$\txnunub{e}$ from $K^\pm_{e3}$ of $\approx1.5\%$, dominated by the
$K^\pm_{e3}$ branching ratio uncertainty.  Small detector calibration
uncertainties, $0.5\%$ on the calorimeter and muon toroid energy
scale, affect the measured $\txnunub{\mu}$ flux and also contribute
substantial uncertainties to both the muon and electron neutrino
fluxes.  Sources of $\txnunub{e}$ other than $K^\pm$ decay have larger
uncertainties, at the 10--20\% level, because of the lack of a direct
constraint from the data.

\begin{figure}[tbp]
\epsfxsize=\textwidth
\epsfbox{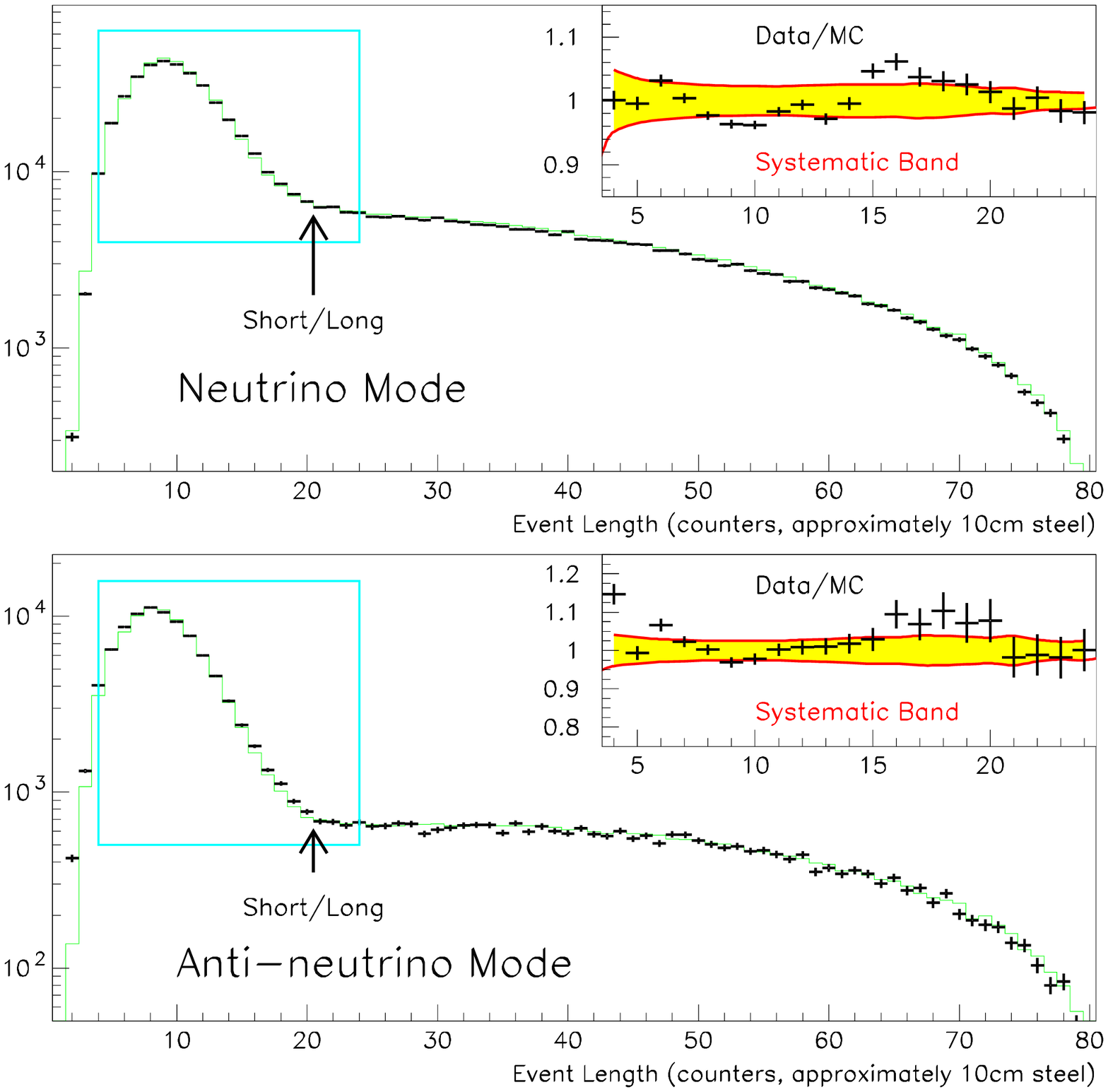}
\caption{Length distributions in the data from the neutrino
and anti-neutrino beams.  Neutral-current/charged-current separation
is made at a length of $20$~counters, approximately $2.1$~m of steel.}
\label{fig:length}
\end{figure}

\compressedsection{Extraction of \stw}

Events selected for this analysis are required to deposit at least
$20$~GeV in the target calorimeter to ensure efficient triggering and
vertex identification.  The location of the neutrino interaction must
be within the central $2/3^{\rms rds}$ of the calorimeter's transverse
dimensions, at least $0.4$~m of steel-equivalent from the upstream end
of the calorimeter, and at least $2.4$~m from the downstream end.  The
first requirement reduces the misidentification of $\txnunub{\mu}$
events with muons exiting the side of the calorimeter; the second
reduces non-neutrino backgrounds, and the third ensures sufficient
calorimeter downstream of the interaction to measure the event length.
Small backgrounds from cosmic-ray and muon induced events are
subtracted from the sample.  After all cuts, $1.3$~million and
$0.30$~million events are observed in the neutrino and anti-neutrino
beam, respectively.  The ratios of neutral-current candidates (short
events) to charged-current candidates (long events), \rmeas{}, are
$0.4198\pm0.0008$ in the neutrino beam and $0.4215\pm0.0017$ in the
anti-neutrino beam.

\rmeas{}\ is related to the ratios of cross-sections and \stw\ using a
detailed detector and cross-section Monte Carlo simulation with the
tuned flux (Figure~\ref{fig:flux-tune}) as input.  This
Monte Carlo must predict the substantial cross-talk between the samples.
In the neutral-current sample, the backgrounds in the neutrino and
anti-neutrino beam from $\txnunub{\mu}$ charged-current events are
$19.3\%$ and $7.4\%$, and the backgrounds from 
$\txnunub{e}$ charged-currents are $5.3\%$ and $5.8\%$.  The
charged-current sample has only a $0.3\%$ background from
neutral-current events for each beam.

The important details of the detector for this analysis are the
calorimeter response to muons, the measurement of the neutrino interaction
vertex, and the range of hadronic showers in the calorimeter.  The
efficiency, noise and active areas of the scintillation counters are
all measured using neutrino data or muons from the testbeam.
Longitudinal and transverse vertex resolutions and biases are studied
using a GEANT-based detector Monte Carlo.  The longitudinal bias,
which directly impacts the length measurement, is measured from the
data using track-based vertices in events with two energetic final
state muons.  Hadronic shower length in the calorimeter is measured
using hadrons from the testbeam.  To study possible effects from the
difference in strange-quark content between neutrino-induced and
$\pi^-$-induced showers, hadronic showers from $K^-$s are used as a
cross-check.  No significant differences are observed.  Measured
detector parameters are varied within their uncertainties in
the Monte Carlo to study systematic errors associated with this
simulation.

The cross-section model is of paramount importance to this analysis.
Neutrino-quark deep-inelastic scattering processes are simulated using
a leading-order cross-section model.  Neutrino-electron scattering and
quasi-elastic scattering are also included.  Leading-order parton
momentum distributions come from a modified Buras-Gaemers
parameterization\cite{BGpar} of structure function data from the CCFR
experiment\cite{Seligman} which used the same target-calorimeter and
cross-section model as NuTeV.  The parton distributions are modified
to produce $u$ and $d$ valence and sea quark asymmetries consistent
with muon scattering\cite{new-NMC} and Drell-Yan\cite{NuSea} data.
The shape and magnitude of the strange sea come from an analysis of
events in CCFR with two oppositely charged muons (e.g., $\nu q\to\mu^-
c$, $c\to\mu^+ X$)\cite{SAR}.  Mass suppression from heavy quark
production is generated in a slow-rescaling model whose parameters are
measured from the same dimuon data.  The charm sea is taken from the
CTEQ4L parton distribution functions\cite{CTEQ}.  The magnitude of the
charm sea is assigned a $100\%$ uncertainty and the slow-rescaling
mass for $(\txnunub{} )c\to(\txnunub{} )c$ is varied from $m_c$ to
$2m_c$.  Our parameterization of $R_{\rms long}=\sigma_L/\sigma_T$ is
based on QCD predictions and data\cite{Whitlow} and is varied by
$15\%$ of itself in order to estimate uncertainties.  Electroweak and
pure QED radiative corrections to the scattering cross-sections are
applied using computer code supplied by Bardin\cite{Bardin}, and
uncertainties are estimated by varying parameters of these
corrections.  Possible higher-twist corrections are considered with a
$100\%$ uncertainty using a VMD-based model which is constrained by
lepto-production data\cite{Pumplin}.

The key test of the Monte Carlo is its ability to predict the length
distribution of events in the detector.  Figure~\ref{fig:length} shows
good agreement between the data and Monte Carlo within the systematic 
uncertainties.

\begin{table}[tbp]
\begin{center}
\begin{tabular}{|r|c|}
\hline
  SOURCE OF UNCERTAINTY & $\delta\sin^2\theta_W$ \\
\hline
  { {\em Statistics}: \hfill \Red{Data} } & \Red{\bf{0.00188}}  \\
  { Monte Carlo } & {0.00028} \\ \hline
  { \bf TOTAL STATISTICS \hfill } & {0.00190} \\ \hline\hline
  { $\nu_e/\nub_e$  } & {0.00045} \\
%  { {\em $\nu_e/\nub_e$ Flux}: \hfill $K^\pm$ ($1.4\%$) } & {0.00029} \\
%  { $K_L$, Charm ($20\%$), $\mu$ ($10\%$) } & {0.00034} \\
  { Energy Measurement} & {0.00051} \\
%  { {\em Energy Measurement}: \hfill  Calibrations ($0.5\%$)} & {0.00049} \\
%  { Muon Energy Deposition ($3\%$)} & {0.00012} \\
%  { Energy Resolution } & {0.00006} \\
  { Event Length} & {0.00036} \\
%  { {\em Event Length}:  \hfill Hadron Shower } & {0.00022} \\
%  { Longitudinal Vertex Determination } & {0.00012} \\
%  { Counter Edge Location } & {0.00024} \\
%  { Counter Efficiency and Noise } & {0.00011} \\
%  { {\em Misc.}:  \hfill Transverse Vertex } & {0.00003} \\
  \hline
  { \bf TOTAL EXP. SYST. \hfill } & {0.00078} \\ \hline\hline
  { Radiative Corrections } & {0.00051} \\
  { Strange/Charm Sea } & {0.00036}  \\
%  { {\em Sea Quarks}: \hfill Strange Sea } & {0.00035}  \\
%  { Charm Sea } & {0.00002} \\
%  { $V_{cd}$ } & {0.00006} \\
  { Charm Mass } & {0.00009}  \\
  { $u/d$, $\ubar/\dbar$ } & {0.00027}  \\
%  { {\em Other $\nu/\nub$ Cross-Section Differences}: \hfill 
%       $\sigma^\nu/\sigma^{\nub}$ } & {0.00021}  \\
%  { Non-Isoscalar Target } & {0.00017} \\
  { Longitudinal Structure Function } & {0.00004}  \\ 
  { Higher Twist } & {0.00011} \\ \hline
  { \bf TOTAL PHYSICS MODEL \hfill } & {0.00070} \\  \hline\hline  
  { \bf TOTAL UNCERTAINTY \hfill } & {0.0022} \\  \hline
\end{tabular}
\end{center}
\caption{Uncertainties in \stw}
\label{tab:unc}
\end{table}
\begin{figure}[tbp]
\epsfxsize=\textwidth\epsfbox{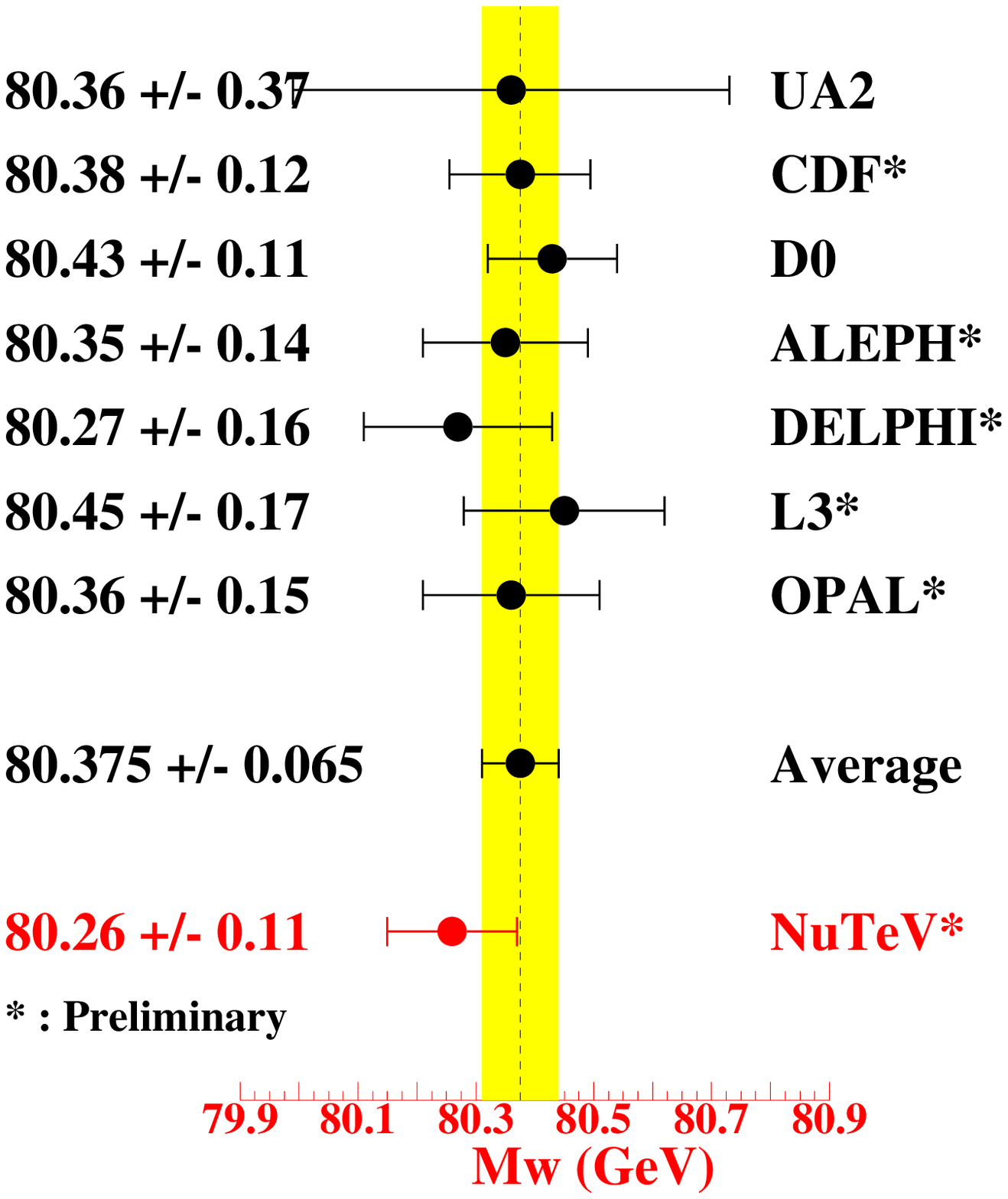}
\parbox{\textwidth}{
\caption{Current direct $M_W$ measurements compared with this result}
\label{fig:world-mw}}
\end{figure}

To compute \stw, a linear combination of 
\rmeas{\nu}\ and \rmeas{\nub}\ was formed,
\begin{equation}
\rmeasm^-\equiv\rmeasm^\nu-\alpha\rmeasm^{\nub},
\end{equation}
where $\alpha$ is calculated using the Monte Carlo such that \rmeas{-}
is insensitive to small changes in the slow-rescaling parameters for
charm production.  $\alpha=0.5136$ for this measurement.  This technique is
similar to an explicit calculation of $R^-$, but here the background
subtractions, the cross-section corrections to Eqn.~\ref{eqn:rminus},
and the dependence on \stw\ are calculated by Monte Carlo.  This
approach explicitly minimizes uncertainties related to the suppression
of charm production, largely eliminates uncertainties related to
scattering from sea quarks, and reduces many of the detector
uncertainties common to both the $\nu$ and $\nub$ samples.
Uncertainties in this measurement of \stw\ are shown in
Table~\ref{tab:unc}.

The preliminary result from the NuTeV data is
{\ifnum\FOOTNOTE=0 \else
\footnote{The weak radiative correction applied to extract \stw$^{\rms
on-shell}$ from the measured quantities has changed since
the presentation at Moriond due to an error in the implementation of
the Bardin code for radiative corrections.  Two other small
experimental corrections, for muon energy deposition and for charm
semi-leptonic decays, were improved as well.  The net shift in the
result, $0.0054$, is dominated by the fix in the implementation
of the radiative corrections.}\fi}
\bea
\stw^{({\rms on-shell)}}&=&0.2253\pm0.0019({\rmt stat})\pm0.0010({\rmt syst})
\nonumber \\ &&
-0.00142\times\left( \frac{\mtop^2-(175~{\rm\textstyle GeV}
)^2)}{(100~{\rm\textstyle GeV})^2}\right ) \nonumber \\ &&
+0.00048\times\log_e\left( \frac{\mhiggs}{150~{\rm\textstyle GeV}}\right).
\eea
The small residual dependence of our result on \mt\ and \mhiggs\ comes
from the leading terms in the electroweak radiative 
corrections\cite{Bardin}.  Since 
\stw$^{({\rms on-shell)}}\equiv1-\mwmw/\mzmz$, this result is equivalent
to
\bea
\mw&=&80.26\pm0.10({\rmt stat})\pm0.05({\rmt syst})
\nonumber \\  &&
+0.073\times\left( \frac{\mtop^2-(175~{\rm\textstyle GeV}
)^2)}{(100~{\rm\textstyle GeV})^2}\right ) \nonumber \\ &&
-0.025\times\log_e\left( \frac{\mhiggs}{150~{\rm\textstyle GeV}}\right).
\eea
A comparison of this result with direct measurements of \mw\ is shown
in Figure~\ref{fig:world-mw}.

It is possible to extract the NuTeV result in a model-independent
framework, where the result is expressed in terms of combinations of
the left and right-handed quark couplings.  The linearized constraint
(expanded around one-loop couplings at an average
$\log_{10}\left( \frac{-q^2}{1~{\rms GeV}^2}\right) \approx1$ for NuTeV's 
central value of \stw) is
\bea
0.4530-\stw~&=&~0.2277\pm0.0022~ \nonumber \\ 
&=&~0.8587u_L^2+0.8828d_L^2-1.1657u_R^2
-1.2288d_R^2.
\eea
Note the similarity of this result to $1/2-\stw=g_L^2-g_R^2$, the definition
of the Paschos-Wolfenstein $R^-$ in Eqn.~\ref{eqn:rminus}.

(It is also possible to combine the NuTeV result with data from NuTeV's 
predecessor, the CCFR experiment.  Adding the CCFR data\cite{CCFR} in the
\rmeas{-}-based method described above, we obtain a slight improvement
in precision, 
\stw$=0.2255\pm0.0018({\rmt stat})\pm0.0010({\rmt syst})$.)

\compressedsection{Conclusions}

The NuTeV experiment has completed its data-taking and has extracted a
preliminary result for \stw$^{\rms(on-shell)}$, which is equivalent to
\mw\ in the Standard Model.  The precision of this result is approximately
a factor of two improvement over previous measurements in \nunuc\ scattering
because of the reduced systematics associated with measuring the 
Paschos-Wolfenstein ratio, $R^{-}$.  This result is consistent with
the average of direct \mw\ data.

{\em We would like to gratefully acknowledge the substantial
contributions in the construction and operation of the NuTeV beamlines
and the refurbishment of the NuTeV detector from the staff of the
Fermilab Beams and Particle Physics Divisions.
%This work was supported in
%part by funds from the United States Department of Energy and National
%Science Foundation and by the Alfred P. Sloan Foundation.
}

%\vspace{-.15in} 


\begin{thebibliography}{99}
\small
\linespace{1.0} 
\vspace{-1.5ex}
\bibitem{langacker} P.~Langacker, {\em et al.\,}, \rmp{64}{1991}{87} .\vspace{-2ex}
\bibitem{oscpaper} K.S.~McFarland, D.~Naples, {\em et al.\,}, 
      \prl{75}{1995}{3993} .\vspace{-2ex}
\bibitem{llewellyn} C.H.~Llewellyn Smith, \np{B228}{1983}{205} .\vspace{-2ex}
\bibitem{CCFR} K.S.~McFarland, {\em et al.\,}, \epj{C1}{198}{509} .\vspace{-2ex}
\bibitem{CDHS} A,~Blondel, {\em et al.\,}, \zp{C45}{1990}{361} . \vspace{-2ex}
\bibitem{CHARM} J. Allaby, {\it et al}., \zp{C36}{1985}{611} .\vspace{-2ex}
\bibitem{Paschos-Wolfenstein} E.A.~Paschos and L.~Wolfenstein, 
               \prev{D7}{1973}{91} .\vspace{-2ex}
\bibitem{BGpar}  A.J. Buras and K.J.F. Gaemers, \np{B132}{1978}{249} .\vspace{-2ex}
\bibitem{Seligman} W.G.~Seligman, {\em et al\,}, 
    \prl{79}{1997}{1213} .\vspace{-2ex}
\bibitem{new-NMC} M.~Arneodo, {\em et al.\,}, \np{B487}{1997}{3} .\vspace{-2ex}
\bibitem{NuSea} E.A.~Hawker, {\em et al.\,}, \prl{80}{1998}{3715} .\vspace{-2ex}
\bibitem{SAR} S.A.~Rabinowitz, {\em et al.\,}, \prl{70}{1993}{134} .\vspace{-2ex}
\bibitem{CTEQ} CTEQ Collaboration, \prev{D55}{1997}{1280} .\vspace{-2ex}
\bibitem{Whitlow}
    L.W. Whitlow, {\em SLAC-Report}-357, 109 (1990) .\vspace{-2ex}
\bibitem{Bardin} D.Yu.~Bardin, V.A.~Dokuchaeva,
        JINR-E2-86-260 (1986); and private communication.\vspace{-2ex}
\bibitem{Pumplin} J. Pumplin, \prl{64}{1990}{2751} .
     $S_0\leq2$~GeV$^2$ is allowed by data summarized in 
     M.~Virchaux and A.~Milsztajn, \pl{B274}{1992}{221} .\vspace{-2ex}

\end{thebibliography}
\end{document}